\documentclass[preprint,aps,amsmath]{revtex4}
\usepackage{graphicx}

\begin{document} 

\title{ZM theory IV: Introduction to quantum concepts, Klein-Gordon and Dirac's equations}

\author{Yaneer Bar-Yam}

\affiliation{New England Complex Systems Institute \\ 24 Mt. Auburn St., Cambridge, Massachusetts 02138}

\begin{abstract}
We describe a general approach to the correspondence of ZM theory with quantum electrodynamics. As a first step, we show the correspondence of helical clock-field states with plane wave states of the Dirac equation. Specifically, defining the direction of time as the gradient of the field in the combined field and space dimensions, for constant gradients, results in states that are consistent with conventional plane wave Dirac equation eigenfunctions. Particles and antiparticles, as well as up and down spins are related by axis inversions. The Dirac wavefunction represents the clock-field and the observer defined direction of time and spatial coordinate axes rotation relative to the region of observation. Additional steps showing the correspondence of ZM theory to quantum mechanics and the Dirac equation are deferred to subsequent papers.
\end{abstract}


\maketitle

\section{Overview}

In this paper we continue an investigation of a novel assumption about the relationship of fields, space and time.\cite{ZM1,ZM2,ZM3} The central assumption is that time can be obtained from the values of a field that at any location has the behavior of a cyclic variable (similar to an analog clock). The observer's time is found from the spatial variation of the field as well as its cyclical progression. This treatment of the field variable $Z$ over the space manifold $M$ leads to ZM theory. This paper is largely devoted to introducing the basic approach to a correspondence with quantum theory and quantum electrodynamics, introducing the differential equation strategy that is compatible with the Klein-Gordon equation (and in the non-relativistic limit Schroedinger's equation),  and showing that the case of linear spatial variation of the ZM clock-field can be mapped onto the plane wave eigenstates of Dirac's equation. For completeness we also generalize previous papers by including the extension of the incremental treatment (corresponding to classical physics by analytic continuation) to more than one dimension of the manifold. 

\section{Toward Quantum Electrodynamics} 

In the development of ZM theory we must consider each step carefully due to differences of the underlying formalism from conventional treatments. It may, however, be helpful to describe an overall perspective framing the correspondence of ZM theory with quantum electrodynamics in anticipation of the more detailed treatments of each aspect. 

Our development of ZM theory has as its objective describing classes of possible states of a clock-field (or multiple clock-fields) as they embody the observer-system relationship. Starting from a clock-field that does not vary in space, we introduce smooth variations in the clock field. As a general taxonomy there is a natural distinction between two characteristic classes of clock-field spatial variations:
\begin{itemize}
\item A linear variation of the phase, corresponding to a helical variation of the clock-field over space. We associate these states to solutions of Dirac's equation.
\item An oscillatory variation of the phase that is small compared to the cycle scale, corresponding to fluctuations of the clock-field over space. We associate these states to solutions of Maxwell's equations.
\end{itemize}
These behaviors are natural elementary excitations of the clock-field and natural names for them would be helices and ripples. Thus, the correspondence with electrodynamics will be described primarily through showing the mathematical correspondence of helices and ripples to electron and photon degrees of freedom in quantum electrodynamics. It is appealing that incrementally (locally), there is no difference between the two types of excitations allowing one to be transformed to the other (e.g. by Gauge transformations). The difference between the excitations arises over distances comparable to the natural length scale of the theory---the size of the proper-time cycle given by the inverse mass, the Compton wavelength. 

There are a number of steps to describing the mathematical correspondence of the ZM theory excitations and electrodynamics. These include describing the correspondence of the states themselves, the coupling of helices and ripples, interactions between multiple excitations including statistics, and the nature of measurement. 
The clock-fields of helices and ripples require different equations in large part due to the distinct geometry of observer time, which is fixed over space-time for the helices and varies for the ripples. A fixed geometry of time leads to pure helical excitations being described by the Dirac equation, and the oscillatory geometry of time for ripples leads them to be described by homogeneous Maxwell's (wave) equations. We can then consider a helical clock-field with ripple variations, as well as the ripple clock field with helical aspects to obtain the electromagnetic coupling in the Dirac equation, and sources in the electromagnetic field equations. Superposition (Boson statistics) of the ripple states follows from the nature of the incremental variation of the clock state, whereas similar superposition does not hold for the helical excitations, whose quantum state superposition and Fermion statistics is more subtle. Measurement theory will not be introduced by the usual Copenhagen interpretation and will have some conceptual modifications. These properties will become manifest as the mathematical development proceeds. Beyond the conventional treatment of electrodynamics, we can also contemplate solving directly in a non-perturbative fashion the ZM clock-field equations and compare with the perturbative treatments in field theoretic electrodynamics. 

While this is a roadmap to the correspondence of ZM theory to electrodynamics, the first step taken in this paper is modest. We explain the mapping of ZM theory representation onto the Dirac equation eigenstates. We will find that even in this modest step there are subtleties to be explained.

\section{Introduction including a multidimensional manifold} 

We review the basic framing of ZM theory from the first paper in the series, ZM1\cite{ZM1}  with sufficient detail for this paper to be essentially self-contained. The original treatment will be augmented to allow for a multidimensional manifold.

We consider a system, $Z$, with some set of distinctly labeled states that perform sequential transitions in a cyclic pattern, i.e. an abstract clock (similar to a conventional ``non-digital" clock consisting of a numbered dial, here with a single moving hand). Discreteness of the clock will not enter into the discussion in this paper. The clock states can therefore be extended to cyclical continuum, $U(1)$. 
{\em A-priori} there is no difference between clockwise and counter-clockwise rotation, however, once one is identified it is distinct from the other.
The state change of the clock defines proper time, $\tau$, as defined by the clock. Since the clock is cyclical it is possible to represent the changing state using an oscillator language: 
\begin{equation}
\psi  = \exp ( - im \tau ),
\label{psi} 
\end{equation}
where 
\begin{equation}
m = 2\pi /T
\end{equation}
is the cycle rate in radians, and $T$ is the cycle period. The clock phase is $c = m \tau$ modulo $2\pi$, though this expression is not analytic so that derivatives should be defined in terms of $\psi$. However, locally with proper choice of the location of the discontinuity, or where analytic continuation is valid, derivatives can be defined in terms of $c$.

The units we use ultimately will correspond to taking the speed of light, and reduced Planck's constant, $\hbar$, to be one. This implies that mass, energy and frequency are measured in the same units. The notation is chosen anticipating that $m$ will become the `rest mass' of the clock when it is reinterpreted as a particle. 

To introduce the space manifold, $M$, we consider a multiple dimensional space manifold, where real valued parameters 
\begin{equation}
\begin{array}{ll}
x&=\{x_1,x_2,...\} \\
&=x_1 \hat x_1+x_2 \hat x_2+...
\end{array} 
\end{equation}
are associated with the environment. We will commonly write the set of parameters indexed by $i$, i.e. $x_i$. For convenience, each of the $x_i$ parameters is measured in the same units as $\tau$. Properties of the environment may lead to variation of the clock state with $x$. For all of the discussion below it is sufficient for us to be considering a neighborhood of a particular point of this manifold.  

The assumption of ZM theory is that observer time can be obtained by the rate of change of the clock, $c$, as a local gradient of the clock state in the Euclidean space given by  $\{\hat \tau,\hat x_1,\hat x_2,...\}$. The `direction of time' is given by the (unnormalized) vector
\begin{equation}
\begin{array}{ll}
(\partial_{\hat \tau} c, \partial_{\hat x_1} c, \partial_{\hat x_2} c,...) 
&=(m, \partial_{\hat x_1} c, \partial_{\hat x_2} c,...) \\
&=  m \hat \tau  + \sum_i \partial_{\hat x_i} c \hat x_i.
\end{array} 
\label{clock}
\end{equation}
The physical interpretation of the ZM theory assumption about the direction of time is that it corresponds to the direction of the maximum rate of change of the clock in this Euclidean space.
We assume that time measures Euclidean distance along the time direction in the same units as $x$ and $\tau$, so the rate of change of the clock phase in the direction of time is 
\begin{equation}
\omega  = \sqrt {m^2  + \sum_{i} (\partial_{\hat x_i}c)^2 }.
\label{omega}
\end{equation}
To demonstrate this as the rate of change of the clock in the time direction we consider possible directions of time (here normalized to be unit vectors):
\begin{equation}
\hat s =  v_\tau \hat \tau - \sum_i v_i \hat x_i , 
\end{equation}
where $v_i$ and $v_\tau$ are respectively the negative and positive directional cosines of the time direction along the $\hat x_i$ axis and $\hat \tau$ axes respectively. The choice of negative signs for the space direction corresponds to the conventional choice of a positive velocity of the system with respect to the observer for positive values of these variables. 
We determine $v_i$ and $v_\tau$ by maximizing the rate of change of the clock written in terms of the clock phase as
\begin{equation}
\begin{array}{ll}
d_s c &= v_\tau \partial_{\hat \tau} c - \sum_i v_i \partial_{\hat x_i} c  \\
&= v_\tau m - \sum_i v_i \partial_{\hat x_i} c 
\end{array} 
\label{generaltime}
\end{equation}
subject to the constraint:
\begin{equation}
v_\tau^2 +  \sum_i v_i^2=1.
\end{equation}
Using a Lagrange multiplier we maximize
\begin{equation}
 v_\tau m  - \sum_i v_i \partial_{\hat x_i} c  - (\omega'/2) \left(v_\tau^2 + \sum_i v_i^2-1\right) 
\end{equation}
The maximization gives
\begin{equation}
\begin{array}{ll}
\omega' v_\tau &= m  \\
\omega' v_i &= - \partial_{\hat x_i} c
\end{array}
\end{equation}
and the constraint equation, which determines $\omega'$:
\begin{equation}
\omega'^2 = m^2+  \sum_i(\partial_{\hat x_i} c) ^2
\end{equation}
Defining $\omega$ as the maximum value of $d_s c$, i.e. the rate of change of the clock in the direction of time, $d_t c$, from these equations we can obtain the expressions
\begin{equation}
\begin{array}{ll}
\omega' &= \omega \\
v_i &= - \partial_{\hat x_i} c / \omega \\
v_\tau &= \sqrt{1-\sum_i v_i^2}.
\end{array}
\end{equation}
These are the usual relativistic relationships of energy, momentum and velocity with the identification\cite{ZM2} of the energy as 
\begin{equation}
\omega = d_t c, 
\end{equation}
and the momentum,
\begin{equation}
k = -\partial_{\hat x_i} c.
\end{equation}
The physical correspondence of $v_i$ to the velocity is still to be established, and will be below. 

As in ZM1 we first restrict our attention to linear variations of the field in space, and allow analytic continuation to extend the domain of the variable $\tau$. This analytic continuation corresponds to the classical limit considered in ZM2.\cite{ZM2}  For linear variation of the clock field, the entire field is determined (aside from an arbitrary offset) by the fixed rate of variation along each of the directions 
\begin{equation}
\begin{array}{ll}
k&=\{k_1,k_2,...\} \\
&=k_1 \hat x_1+k_2 \hat x_2+...  
\end{array}
\end{equation}
We obtain equations describing the relationship of time to proper time and space for this case. Distances are measured as Euclidean in the space given by the axes $\hat \tau$ and $\hat x_i$. By the definition of the direction of time, and that $\omega$ is the rate of change of the clock along this direction, we have that along the time direction, 
\begin{equation}
\omega t = m \tau.
\label{alongtime}
\end{equation}
Assuming that the observer defines time $t$ to be synchronous along observer position $x$, we can more generally write the equation: 
\begin{equation}
\omega t = m \tau  + \sum_i k_i x_i. 
\label{trelation}
\end{equation}
describing the variation of time as a function of $\tau$ and $x_i$, which are not the same as the coordinate axes due to the variation of the clock along the $\hat x_i$ axis.

We can specify locations in space and the clock-field or time dimensions by their coordinates along the axes 
\begin{equation}
(\tau_0,x_{10},x_{20},...)=\tau_0 \hat \tau + \sum_i x_{i0} \hat x_i .  
\end{equation}
We can then write $t$ and $x$ in terms of $\tau_0$ and $x_{i0}$. All expressions can be directly obtained from the geometry of the angle of time with space. For the case of uniform variation of clock time in space, and assuming time has no dependence on the observer defined space coordinates $x_i$, time is given by
\begin{equation}
t = (\omega/m) \tau_0.
\label{ttau00}
\end{equation}
The direction of time implies a shift in the position of the origin of $x$ toward negative values and a reference location translates to the right, with $x_i$ given by: 
\begin{equation}
\begin{array}{ll}
x_i&=x_{i0}+(k_i/m)\tau_0 \\
&=x_{i0}+(k_i/\omega) t.
\label{translation}
\end{array}
\end{equation}
This identifies the velocity 
\begin{equation}
v_i = k_i / \omega.
\label{velocity}
\end{equation}
and shows the relationship of the direction of time to the velocity. 

For the linear case we have treated, the results need not be considered as an analytic continuation of $\tau$ if we insert expressions for $\tau$ into the cyclical clock description by Eq. (\ref{psi}) for $\psi$. This leads us to the discussion of the quantum formalism. First, however, we consider the possibility of a negative $\omega$.

\section{Negative values of $\omega$} 

The mathematical treatment just given of ZM theory can be readily seen to accommodate negative values of $\omega$ whose existence corresponds to the possibility of a counterclockwise clock phase rotation. Since the clock is assumed to represent a set of distinct, i.e. labeled states, counterclockwise rotation is not the same as clockwise rotation. Motivated by reasonableness of symmetry, both clockwise and counterclockwise rotations should be possible. Specifically, in the relationship of observer and system, it is possible for an observer to define time in the opposite sense as the system. Thus, if there is a circumstance such that the coordinate system of the observer is inverted in time, the direction of time is given by the extremum of the rate of change of the clock, but it is the mathematical minimum rather than the maximum.

In the context of the discussion of traditional relativistic quantum mechanics, the existence of antiparticles can be interpreted as arising through symmetry (or rather lack thereof) of particle states under time inversion.\cite{Feynman} This interpretation appears in the correspondence of ZM theory to quantum electrodynamics discussed below. In advance of the connection to antiparticles, we can consider the mathematical treatment to see how definitions should be generalized. The key issue is that the velocity becomes negative under time inversion, but the momentum, fixed by the spatial variation of $\tau$, does not. This result can be found conceptually, geometrically or analytically.

Conceptually, since the direction of time and the velocity are the same quantities, inverting the direction of time results in inversion of the velocity. The momentum is set by the clock-field variation in space which is fixed when the inversion is performed. 

Geometrically, a positive $k_i$ results in a direction of time that is in the quadrant of $(\tau_0,x_{i0})$ space with positive $\tau_0$ and negative $x_{i0}$ (see figure 1 of ZM1). Inversion of the direction of time results in a negative $\tau_0$ and positive $x_{i0}$. Movement of the origin of coordinates along the direction of time results in a shift in the positive $x_{i0}$ direction, implying a negative velocity. 

Mathematically, fixing $\tau$ over space and therefore $k$, time inversion implies the direction of time is the negative of the gradient of the clock field, and the rate of change of $\tau$ along the time direction is 
\begin{equation}
\omega  = - \sqrt {m^2  + \sum_{i} (\partial_{\hat x_i}c)^2 }.
\end{equation}
In all of the equations (\ref{alongtime},\ref{trelation},\ref{ttau00},\ref{translation},\ref{velocity}), $t$ appears in combination with $\omega$ so that under the transformation $t \rightarrow -t$ and $\omega \rightarrow -\omega$ the equations are unchanged. They are therefore valid as written. In particular, Eq. (\ref{velocity}) 
\begin{equation}
v_i = k_i / \omega.
\end{equation}
is a valid expression for the velocity in terms of the momentum  so that a negative $\omega$ results in a velocity which is the opposite sign of the momentum. 

We note that from the point of view of classical physics one might expect that both velocity and momentum would change sign, given the conventional relationship $k=mv/\sqrt{1-v^2}$, which is equivalent to Eq. (\ref{velocity}) for positive $\omega$. The definition of velocity given here might be better considered as analogous to the classical current. This would take care of the relative signs because the current is the product of charge and velocity, where the charge has the opposite sign for antiparticles. Risking some confusion, we will continue to use the notation and terminology of velocity for this quantity.

\section{Quantum Physics correspondence of ZM theory} 

In order to introduce the quantum regime in ZM theory we consider the ZM clock-field for helical states to be analogous to the quantum wavefunction. The major additional effort is to consider the implications of the definition of time as a function of the space and proper-time dimensions. The essential concept is that the proper time is analogous to the wave function phase.  The time derivatives and space derivatives are related to each other by quantum differential equations including the Dirac and Klein-Gordon (and Schroedinger's) equations. These equations  describe the relationship of the space and time derivatives of the clock state through the relationship of time to proper time and space as given by ZM theory.

To proceed we revisit our definitions of physical quantities. First, we reaffirm that our objective is to develop an understanding of the ZM clock field, described in terms of a complex number phase given by: 
\begin{equation}
\psi = \exp (-i m\tau)
\label{partialwave}
\end{equation}
We identify the quantum physical quantities as operators acting on $\psi$ generalizing the classical treatment of mechanics in ZM2 of operators acting on $c=m\tau$. In order to develop the definitions we note that there is a natural ambiguity in the context of describing a cyclical system as described by Eq.  (\ref{partialwave}). Expressions analogous to those used in quantum mechanics for energy and momentum from a differential operator might be defined in two ways:
\begin{equation}
\psi^\dagger \partial \psi 
\end{equation}
\begin{equation}
\psi^{-1} \partial \psi 
\end{equation}
These two expressions are equivalent for a single partial wave. In order to determine which is useful, we will have to understand how plane waves are superposed. The properties of superposition are generally assumed as postulates of quantum theory. Here, we will consider them as we develop the formalism. In the meantime, however, to avoid confusion it seems better to use the conventional quantum notation of adjoint, and note that we will need to justify this choice when superposition is described.

With this caveat we define physical quantities in parallel to the quantum mechanical definitions and consistent with the description of the ZM theory clock-field as analogous to the quantum wavefunction:
\begin{itemize}
\item rest $mass$ is the rate of change of the state of the clock in the cartesian clock (proper time) dimension, 
\begin{equation}
m  = i \psi^\dagger \partial_{\hat \tau}  \psi
\end{equation}
\item $momentum$, $k$, is the negative of the rate of state change with respect to an environment degree of freedom. 
\begin{equation}
k = - i \psi^\dagger \partial_{\hat x} \psi
\end{equation}
\item $energy$, $\omega$, is the rate of change of system state with respect to observer time, $t$, 
\begin{equation}
\omega  =  i \psi^\dagger d_t \psi
\end{equation}
where the full derivative with respect to time is defined to be the derivative in the time direction
\begin{equation}
d_t = \partial_{\hat t}.
\end{equation}
\end{itemize}
Expressions relating these three physical properties are the subject of the quantum differential equations.

For the case of a ZM helix, a linear variation of the clock-field in space, the discussion in the introduction applies and the clock state can be written in terms of the cartesian coordinate axes as 
\begin{equation}
\psi=\exp (i(\sum_i k_i x_{i0}  - m \tau_0))
\end{equation}
Setting the direction of time to be the maximum rate of change of the clock gives us that
\begin{equation}
\omega^2 = \sum_i k_i^2 + m ^2.
\end{equation}
The clock phase in terms of space and time is given by the expression obtained previously in Eq.  (\ref{trelation}) that can be substituted into the exponent of $\psi$ to give
\begin{equation}
\psi = \exp (i(\sum_i k_i x_i  - \omega t))
\label{planewave}
\end{equation}
It is important to emphasize that these expressions are valid for linear variations of the clock phase.  The expression in Eq. (\ref{planewave}) is immediately recognizable as a plane wave which mathematically corresponds to a conventional momentum and energy eigenstate in classical quantum mechanics. While this is suggestive, a few more steps are needed. At this level of discussion ZM theory proceeds in the reverse direction of conventional quantum theories in that quantum theories assume an equation from which possible physical states are obtained, while ZM theory defines the set of possible states and from this obtains the equations that can describe them. 

Our next objective is to write the plane-wave helix clock-field expressions as differential relationships between the variations of the clock variable in space and time. The usefulness of doing so arises because the linear variation can be generalized to include spatial variations where linear relationships continue to hold incrementally. Candidate differential equations include the conventional ones found in quantum mechanics. The objective is to identify differential equations that apply to as broad a class of clock-fields as possible.

\section{Klein-Gordon and Schroedinger's equation} 

In considering possible differential equations to describe the behavior of the clock field we start by writing the most natural equation in ZM theory. The ZM theory equation for the time derivative of $\psi$ given by Eq. (\ref{generaltime}) in the direction of time
\begin{equation}
d_t \psi = v_{\tau} d_{\hat \tau}\psi  -   \sum_i v_i \partial_{\hat x_i} \psi,
\label{zmeq}
\end{equation}
is complicated because of the implicit relationship of $v_i$ to $k_i$. To avoid this complication we will follow the approach of conventional quantum mechanics which sought differential equations that describe the behavior of quantum systems and have constant coefficients. These equations were historically justified by phenomenological verification. Here, our objective is to determine whether solutions of the equations correspond with the ZM clock field states. We therefore consider differential equations with constant coefficients that can describe clock-field states and the relationship of the derivatives of the clock field to each other, to replicate the equations of quantum electrodynamics. 

Formally speaking, we need not derive the differential equations only show that they are consistent with ZM theory. It is important to recognize that ZM theory does not endorse a particular differential equation except as the differential equation is satisfied for a class of clock-field states and its implicit relationships between time, proper-time and space coordinates. Indeed, we anticipate that no differential equation will be entirely successful as is found from the need for field theoretic treatments of quantum electrodynamics. Still, since the helical clock-field can be described mathematically as a plane wave, any differential equation with plane wave solutions is a candidate.  This includes the conventional quantum mechanical Klein-Gordon and Schroedinger equations. However, it does not include the Dirac equation because it is a matrix equation with Dirac spinor solutions. Therefore, at this level of discussion the correspondence appears trivial for the former but requires explanation for the latter. 

Despite the simplicity for treatment of the Klein-Gordon and Schroedinger equations, it may be helpful to discuss the framework of ZM theory in this context to show how the relationships with these differential equations should be considered.
Thus we can consider the equation 
\begin{equation}
 d_t^2 \psi  = \sum_i \partial _{\hat x_i }^2 \psi  + \partial_{\hat \tau}^2  \psi  
\end{equation}
which is equivalent to the Klein-Gordon equation if we replace the derivative with respect to the proper time by substituting $i \partial_{\hat \tau} \psi $ with the value $m \psi$
\begin{equation}
 d_t^2 \psi  =\sum_i  \partial _{\hat x_i }^2 \psi  - m^2  \psi  
\end{equation}
In considering the utility of this equation we recognize that, since it is not of first order, there will be corrections when $k$ is spatially dependent that may be difficult to describe in this form. The advantage of a first order equation is in describing relatively simply the effect of a spatially varying $k$. Another difficulty with the Klein-Gordon equation is that it does not represent the direction of time, as contained in the velocity in the ZM theory equations. The ability to capture the direction of the observer coordinates will be shown to be a key attribute of the Dirac equation.  

In the non-relativistic limit Schroedinger's equation follows from the use of the expansion of $\omega$ in terms of $k$, resulting in
\begin{equation}
 id_t \psi  \approx  m  \psi -\left(\partial _{\hat x_i }^2/2m \right) \psi 
\end{equation}
There are several differences from the Klein-Gordon equation that make the Schroedinger equation potentially more desirable, in the regime in which it is applicable. The appearance of the first derivative with respect to time suggests that linear approximations will be better behaved. Moreover, the direction of time which relativistically is represented by the velocity, can be approximated to linear order by the momentum which does appear in the equation, though it appears only in second order.

In each of these differential equations we can consider how the equation would be generalized to accommodate the introduction of spatial variation of $k$. This will be addressed in a subsequent paper. Here we focus on the discussion of the Dirac equation eigenstates which require further explanation. The main difference in Dirac's equation compared to Klein-Gordon and Schroedinger's equations is that the structure of the wave function consists of a Dirac spinor. From the point of view of ZM theory Klein-Gordon and Schroedinger's equations do not describe the direction of time, which captures a key aspect of the physical relationship between the observer and the system. We will show that the Dirac equation spinors capture this information in the next section.

\section{Dirac Equation eigenstates} 

A description of the clock-field implies an observer-system relationship. We will show that for ZM helices, the Dirac wavefunction captures both the clock-field as a plane wave, and the direction of time as  a Dirac spinor. The eigenfunctions guarantee the relationship between the helical variation and the direction of time prescribed by ZM theory.

The Dirac equation can be written\cite{Dirac,Huang,BjorkenDrell} as:
\begin{equation}
 id_t \Psi  = -i\sum_i \alpha _i \partial _{x_i } \Psi  + \beta m  \Psi  
\end{equation}
where the $\alpha_i$, and $\beta$ are self-adjoint operators that square to unity and anti-commute with each other, and are generally represented as four by four matrices, with $\Psi$ a four-component vector. 

As a linear equation, all solutions can be written as a superposition of the Dirac equation eigenfunctions. The eigenfunctions can be chosen to be written as a product of a Dirac spinor and a plane wave
\begin{equation}
\Psi  = \chi \psi 
\end{equation}
where 
\begin{equation}
\psi  = \exp (i(\sum_i k_i x_i-\omega t))
\end{equation}
is the plane wave with the same form as the ZM theory helix representation in Eq. (\ref{planewave}). $\chi$ is a four component vector (Dirac spinor) satisfying
\begin{equation}
\omega \chi  = \sum_i \alpha _i k_i \chi  + m  \chi.
\end{equation}
The matrix form of the equation is
\begin{equation}
\omega \chi  = \left( \begin{array}{cccc}
 m & \  0 &\  k_3 & \  k_{-} \\
 0 & \  m & \  k_{+} & -k_3 \\
k_3 & \  k_{-} & -m & \  0 \\
k_{+} & -k_3 & \  0 & -m \end{array} \right)  \chi.
\end{equation}
where $k_{-}=k_1-ik_2$ and $k_{+}=k_1+ik_2$. For a particular value of $k_i$ there are four independent Dirac wavefunctions $\Psi_a$, $a=1,2,3,4$, associated with $\chi_a$, which can be written explicitly as the columns of the matrix 
\begin{equation}
\chi  =\sqrt{\frac{|\omega|+m}{2m}} \left( \begin{array}{cccc}
 1 & 0 & \frac{-k_3}{|\omega|+m} & \frac{-k_{-}}{|\omega|+m} \\
 0 & 1 & \frac{-k_{+}}{|\omega|+m} & \frac{k_3}{|\omega|+m} \\
 \frac{k_3} {|\omega| + m} & \frac{k_{-}}{|\omega| + m} & 1 & 0 \\
 \frac{k_{+}} {|\omega| + m} & \frac{-k_3} {|\omega| + m} & 0 & 1 \end{array} \right),
 \label{chi}
\end{equation}
where arguments about analyticity from the solutions at $k_i=0$ have been used to conclude that the first two columns should be used for $\omega=\sqrt{k^2+m^2}$ and the last two for $\omega=-\sqrt{k^2+m^2}$. Therefore, the two pairs of solutions share the same $k_i$ but not the same plane wave due to the difference in sign of $\omega$. For convenience we introduce the variable $e_a$ which takes the values $\{1,1,-1,-1\}$ when $a$ takes the values $\{1,2,3,4\}$ respectively. 

The expressions for the Dirac eigenfunctions correspond to a normalization chosen in conventional electrodynamics so that the relativistic volume contraction is compensated for by the amplitude square of the Dirac spinor
\begin{equation}
\chi_a^\dagger\chi_a = |\omega|/m.
\end{equation}
Keeping this in mind, the identity of other operators is defined to include the normalization. Specifically, by direct evaluation, $\alpha_i$ is associated with the $i$th component of the velocity. 
\begin{equation}
\begin{array}{ll}
\Psi_a^\dagger \alpha_i \Psi_a &= \chi_a^\dagger \alpha_i \chi_a \\
&= e_a k_i / m \\
&= e_a (k_i/|\omega|) (|\omega| / m) \\
&=  v_i (|\omega| / m)
\end{array}
\end{equation}
and $\beta$ with the `density'
\begin{equation}
\begin{array}{ll}
\Psi_a^\dagger \beta \Psi_a &= \chi_a^\dagger \beta \chi_a \\
&= 1 \\
&= e_a (m/|\omega|) (|\omega| / m) \\
&= v_{\tau} (|\omega| / m)
\end{array}
\end{equation}
which we identify as the $\tau$ directional cosine of the time direction in ZM theory. We have adopted the convention introduced in Section IV on negative values of $\omega$ where the velocity is given by
\begin{equation}
\begin{array}{ll}
v_i & = e_a (k_i / |\omega|) \\
&= k_i / \omega
\end{array}
\end{equation}
Note also that since we have not yet introduced a general formalism of expectation values in ZM theory our identification of these quantities is only through their explicitly calculated values from the eigenfunctions.

In order to make a connection to ZM theory we reinterpret these entities as follows. The Dirac wavefunction is interpreted as a product of a representation of the direction of time, the Dirac spinor $\chi$, and a representation of the clock-field, the plane wave $\psi$. At the first level of discussion of such a mapping we require that the direction of time is fixed in space and time. This is consistent with the eigenvectors of the Dirac equation. Considering only these eigenvectors, we multiply the Dirac equation on the left by $\chi^\dagger$ to obtain 
\begin{equation}
 i\chi^\dagger d_t \chi \psi = -i\sum_i \chi^\dagger \alpha _i \partial _{x_i } \chi \psi  + \chi^\dagger \beta m  \chi \psi  
\end{equation}
since $\chi$ is constant we pass it through the space and time derivatives to obtain
\begin{equation}
 i\chi^\dagger\chi d_t \psi = -i\sum_i \chi^\dagger \alpha _i \chi \partial _{x_i }  \psi  + \chi^\dagger \beta \chi  m \psi  
\end{equation}
Inserting the results above and dividing by the normalization $\chi^\dagger\chi$  gives
\begin{equation}
 i d_t \psi = -i \sum_i v_i  \partial _{x_i }  \psi  + v_{\tau} m \psi  
\end{equation}
Recognizing the right term as given by the partial derivative with respect to proper time we obtain the ZM equation, where the velocity is determined by the direction of time
\begin{equation}
 id_t \psi  = -i \sum_i v _i \partial _{x_i } \psi  + i v_{\tau} \partial _\tau  \psi  
\end{equation}
It is apparent from the derivation that at this point we have only considered the eigenfunctions of the Dirac matrix. However, since all solutions are determined by these eigenfunctions we can defer the complete discussion of solutions to the consideration of superposition. Still, the behavior of ZM theory must be further explained in order to make a full connection with Dirac theory. Among the interesting questions are the appearance of particle and anti-particle as well as spin degrees of freedom. 

\section{antiparticles and spin}

The existence of four eigenvector solutions of the Dirac equation for a particular value of $k$ must still be addressed. These are identified with the four combinations of two spin and particle/anti-particle states. The appearance of antiparticles and spin in Dirac's equation is conventionally remarkable because the framing of the equation is as a description of a structureless point object to which the wavefunction is related by measurement theory expectation values. 

ZM theory has a different framing based upon the observer's definition of space-time as described in the fourth comment of the introduction in ZM1. According to this interpretation, the definition of space time $(t,x)$ is that of an observer in one region of space (region $A$) extrapolating a local definition of time and space to another region of space (region $B$). The space-time properties in region $B$ can then be identified through the behavior of the clock field.  Such an extrapolation implies that underlying the Dirac equation is a relationship between two different coordinate systems. This relationship has more structure than a point object and gives reason for the existence of particle, antiparticle, and spin states. 

Specifically, if we assume the possibility of axis inversions in the relationship between different observer coordinate systems, i.e. coordinates associated with different regions of space, there are four distinct states. These distinct states are defined by inversion of the direction of time and inversion of a spatial axis. The former is a discrete transformation that cannot be reached by continuous transformation of the coordinate systems because it changes a forward moving clock into a backward moving clock. The latter is discrete because it results in parity inversion of the relationship of the axes, and continuous transformations of coordinates from one handedness to the other are not possible.

Among the axis inversions that are possible, the most conceptually natural one would be one that would invert the axis along the velocity vector. However, such an inversion would result in a sign change of the momentum and velocity, not because they have been changed but because we are measuring them in the opposite direction. Since this takes us out of the subspace of states we are considering, we would have to perform additional transformations, to make the connection mathematically. 

An alternative approach that makes clear the connection between axis inversion and spin without the same difficulty, is to invert the axes that are perpendicular to the velocity. Our argument identifies the handedness of the coordinate system with the change of spin, hence inverting these axes should change the spin and does not impact on the direction of the velocity. We verify the effect of perpendicular axis inversion for momentum and velocity along the $\hat x_3$ direction and leave other verifications to known Dirac equation symmetries or exercises.

A one axis inversion operator when applied to a Dirac spinor leads to the inversion of that component of the velocity while leaving others unchanged. Specifically, let $\iota_i$ be the $i$th axis inversion operator. Then taking a Dirac spinor $\chi$ 
\begin{equation}
\chi' = \iota_i \chi
\end{equation}
we have for $i' \ne i$
\begin{equation}
\chi'^\dagger \alpha_{i'} \chi' =  \chi^\dagger \alpha_{i'} \chi 
\end{equation}
and
\begin{equation}
\chi'^\dagger \alpha_i \chi' =  - \chi^\dagger \alpha_i \chi 
\end{equation}
We note that $\iota_i$ can be multiplied by any complex number of unit magnitude. We can interpret this flexibility as arising because any complex number phase can be absorbed into a fixed offset of the plane wave representation of the clock-field and is not part of the Dirac spinor representation of the velocity. Alternatively, it corresponds to a translation of the coordinate system. This translation results in a rotation of the clock-field phase by an amount smaller than the clock cycle.

The anti-commuting properties of $\beta$ and  $\alpha_i$ imply that they invert three out of four of the four dimensional Euclidean $( \hat \tau, \hat x_i )$ space, each leaving its corresponding axis uninverted. This implies that we can construct a one axis inversion operator as a product of the three operators that do not correspond to that axis. Thus, choosing a particular phase for convenience:
\begin{equation}
\begin{array}{ll}
 \iota_1 &= -i \alpha_2 \alpha_3 \beta \\
& = \sigma_1 \beta
\end{array}
\end{equation}
where $\sigma_1$ is a matrix whose non-zero elements are contained in two block diagonal 2x2 matrices that are the first Pauli matrix. Explicitly
\begin{equation}
 \iota_1  =   \left( \begin{array}{cccc}
\  0 & \  1 & \  0 & \  0 \\
\  1 & \  0 & \  0 & \  0 \\
\  0 & \  0 & \  0 & -1 \\
\  0 & \  0 & -1 & \  0 \end{array} \right).
\end{equation}
Similarly,
\begin{equation}
\begin{array}{ll}
 \iota_2 &= -i \alpha_3 \alpha_1 \beta \\
& = \sigma_2 \beta
\end{array}
\end{equation}
where $\sigma_2$ is constructed similarly to $\sigma_1$ but from the second Pauli matrix. Explicitly
\begin{equation}
 \iota_2  =   \left( \begin{array}{cccc}
\   0 & -i & \  0 & \  0 \\
\  i & \  0 & \  0 & \  0 \\
\  0 & \  0 & \  0 &\   i \\
\  0 & \  0 & -i & \  0 \end{array} \right).
\end{equation}
Since an overall phase does not matter, we can also multiply by $i$ and choose $\iota_2$ to be real. 

We consider the case of velocity and therefore momentum along the $\hat x_3$ direction so that $k_i = 0$ for $i=1,2$. We then have a Dirac spinor given by columns of the matrix:
\begin{equation}
\chi  =\sqrt{\frac{|\omega|+m}{2m}} \left( \begin{array}{cccc}
 1 & 0 & \frac{-k_3}{|\omega|+m} & 0 \\
 0 & 1 & 0 & \frac{k_3}{|\omega|+m} \\
 \frac{k_3} {|\omega| + m} & 0 & 1 & 0 \\
0 & \frac{-k_3} {|\omega| + m} & 0 & 1 \end{array} \right).
\end{equation}
By direct multiplication it is possible to check that applying $\iota_1$ or $\iota_2$ switches columns 1 and 2 and columns 3 and 4. Overall multiplication by a phase can be cancelled or could have been avoided by defining differently the transformation operators.

The description of the use of axis inversion to change the spin is not a complete description of the mapping of the ZM theory onto the representation of the Dirac equation. In the next section we describe how we can include rotations.

\section{Three dimensional coordinate transformations}

The discussion of spin as arising from the anti-alignment of coordinate axes of observer with the observed region of space suggests that we should more generally allow and specify spatial coordinate transformations, i.e. rotations, between the observer and the observed region of space. Strictly this identification is not necessary at this point in the development of the theory since our objective of showing that the Dirac eigenstates satisfy the ZM theory equations has been achieved, however it provides a possible interpretation of the Dirac spinor notation consistent with ZM theory. This interpretation can guide further development of the theory.

Mathematically, the introduction of rotations can be understood quite analogously to the conventional discussion of the transformation of the Dirac equation under rotation. The Dirac spinor is a faithful representation of the Lorentz group, including rotations and boosts, though it is doubled enabling the representation of inversions. While we have demonstrated that ZM theory is Lorentz covariant,\cite{ZM1} we have chosen to separate boosts from rotations in discussing the correspondence of the Dirac equation with ZM theory. We have reinterpreted boosts as the tipping of the time direction, where the angle of time is the velocity. The introduction of rotations can be understood much more closely to their conventional discussion. 

To proceed, we generalize the interpretation of the Dirac spinor from a representation of the direction of time (velocity) to include coordinate rotations of the observer frame of reference. 
We note that the Dirac spinor has more degrees of freedom (six, excluding the overall phase and normalization from the four complex components) than the direction of time (three degrees of freedom). This is explained by allowing the spinor to represent spatial rotations (three degrees of freedom). The transformation
\begin{equation}
\chi' = \exp ( i \sum_i \theta_i \sigma_i) \chi
\end{equation}
provides the rotation around the axis in the direction of the vector $\theta_i$ by the angle given by the magnitude of this vector. As before, $\sigma_i$ are the four by four matrices constructed from the corresponding Pauli matrices as block diagonals. Thus, the set of Dirac spinors can represent the possible observer coordinate system rotations. 

The conventional interpretation of these rotations begins from the assumption that the wavefunction is a representation of the state of the system, i.e. particle. The wavefunction, being an extended entity in space, is not invariant under rotations. Rotations transform the state of the system in a way which would be equivalent to changes in observer coordinate system orientation with respect to the system.  The correspondence of Dirac's equation with ZM theory starts differently by interpreting the Dirac spinor as representing the relationship of observer and system. Just as with the direction of time, the relationship between the observer and the system is determined by the process of extrapolation of the coordinate system. The extrapolation may involve coordinate rotations. These rotations are captured in the Dirac spinor. Unlike the direction of time, however, the rotation of the coordinate system cannot be directly inferred from the clock field at the location of observation. Instead it must be obtained from the non-local structure of the clock field. Hence it is convenient to have a local system representation such as the Dirac equation that allows all possible coordinate rotations. 

As it is conventionally shown, a rotation of the spinor by $\theta_i$ and a rotation of the momentum vector by the same vector gives rise to another eigenstate of the system. In the Dirac notation, such a state is not represented the same as that generated by imposing initially the rotated velocity on one or the other of the spin states. This is conventionally understood as due to the rotation of the spin as well as the velocity, where intermediate spin orientations can be obtained by superposition of two spin states. In ZM theory the lack of equivalence of the rotated state with other states of the same momentum is significant in that the local direction of the momentum and velocity is independent of the rotation of the coordinate system and both must be specified by the representation system.

\section{notes}

There are many notes that may be relevant to our discussion of ZM theory and the Dirac equation. They clarify some conceptual issues as well as point to additional aspects of the Dirac equation, quantum mechanics and quantum electrodynamics that are still to be developed.

\subsection{Axis inversions}

First, we note that the existence of symmetries under transformations of time and space that relate distinct particle states, such as particle/antiparticle and spin, is today a fundamental theoretical concept. Changes among these distinct states are associated with time and space inversion, a kind of broken symmetry, or choice of symmetry group representation. The association of symmetry with such particle states might be said to have been discovered historically through the Dirac equation, since the experimental observation of antiparticles occurred afterwards, and while spin was known beforehand, it is required by the mathematics of the Dirac equation, even when it is not assumed. Conceptually, the discovery arises because the Dirac equation is conventionally assumed to describe the properties of a point particle. For a point particle, there seems no need to assume a such a change of state under time and and space inversion, only a change of direction of motion; a state change requires some form of ``internal" degree of freedom. The Dirac equation implies that we need such internal degrees of freedom to characterize the spin and particle/antiparticle states. In contrast, in ZM theory the framing of the theory in terms of coordinate system relationships implies a change of state under transformation is a more natural starting point. The imposition of a symmetry property of coordinate system transformation can therefore be argued to be the reason for the existence of the different particle states. One manifestation of this difference in approach is the conventional interpretation of antiparticle states as ``negative energy states" that must be filled. In contrast, in ZM theory antiparticles arise naturally through the possibility of an observer measuring time counterclockwise. This is consistent with Feynman's interpretation\cite{Feynman} of antiparticles as particles moving backwards in time. 

Second, more specifically, the existence of spin associated with space axis inversion operators implies that the coordinate axes are distinguishable under inversion. This is similar to the distinction of clock states that allows clockwise and counterclockwise to be distinguished. This issue will be further discussed in later papers.

Third, the explanation of the existence of antiparticles and spin degrees of freedom based upon axis inversion requires a second theoretical step, which is the identification of the mechanism by which axis inversion occurs during extrapolation of the space-time axes. Such a demonstration would provide self-consistency in the formulation of ZM theory. This step is left to subsequent papers considering more general variations of clock fields.

\subsection{Axis rotations}

Fourth, the discussion of rotation operators in relation to the Dirac equation and the spinor representation suggests a more axiomatic approach to the correspondence of ZM theory with the Dirac equation. The approach would start by introducing the existence of multiple dimensions associated with the environment and develop a description of the observer-system relationships based upon coordinate transformations, including rotations, translations and inversions. In general, when considering an abstract manifold of environment coordinates, we can consider the existence of multiple dimensions without any direct relationship between them. However, we are particularly interested in the case where the coordinates can be related to each other by rotation through intermediate directions. If the dimensions are related analytically by rotation, coordinate rotations can be different from rotations of the momentum / clock-field spatial variation, which specifies the direction of time associated with the helix. This distinction arises because the direction of time is not orthogonal to the direction of space. Considering these issues naturally suggests a representation that allows the independent description of the clock-field, from the coordinate transformation. This is consistent with our interpretation of the Dirac wavefunction as a product of the representation of the clock-field and the coordinate system relationship, including the observer direction of time.

Fifth, the existence of rotations of the coordinate system independent of the direction of time does not mean, at this point in the discussion, that the direction of time can actually point in a different direction than the direction specified by $k$. Rather, it means that the equation describing the analytic continuation of the solutions may contain such a situation. The existence of such a representation is, however, relevant to the discussion of superposition as well as the generalization of Dirac's equation to include ripples (electromagnetic fields) which will allow the direction of time to differ from the helix determined value of $k$.

Sixth, our discussion provides a geometric interpretation of the correspondence of ZM theory and quantum electrodynamics. A geometric interpretation using a representation of the Lorenz transformation in the language of the Geometric Algebra may also be useful.\cite{Hestenes} In ZM theory we could consider such a geometric interpretation for the four dimensional Euclidean coordinate system formed by the proper-time axis and space axes.

\subsection{Distinct states}

Seventh, one of the interesting issues that are present in quantum mechanics is the difference between the space of possible wavefunctions and the counting of distinct states. Quantum mechanics does not follow the intuition that different system states (wavefunctions) should be counted as different for all purposes. There are many different wavefunctions that are possible system states, which can be formed by superposition of the eigenfunctions. However, physically distinct states are identified by a basis set of eigenfunctions. Alternatively, the counting of physically distinct states is given by the volume of phase space divided by the reduced Planck's constant, a semiclassical concept. While we will not consider this issue fully here, we see that ZM theory provides a conceptual path toward understanding this issue through the distinction between system state and observer-system relationship. According to ZM theory correspondence, both are contained in the quantum wavefunction. Thus, the set of possible wavefunctions includes the set of possible observer-system relationships, not only the set of possible states of the system. In the context of the Dirac wavefunctions, an example of this distinction occurs in the superposition of up and down spins under rotation giving many more wavefunctions than the number of distinct states.

\subsection{Superposition}

Eighth, quantum superposition as a fundamental property of quantum mechanical wave equations, applicable to the Dirac equation, has yet to be shown for ZM theory. While the superposition of multiple solutions of the equation is trivial mathematically, the physical meaning is important. One instance of conventional superposition that has appeared is the superposition associated with rotations. However, such superpositions preserve the norm of the wavefunction and their interpretation in ZM theory as reflecting the relationship between observer and system by coordinate transformation cannot be trivially extended to arbitrary superposition. The more general form of superposition has been alluded to in the context of discussions of the semiclassical approximation in ZM3\cite{ZM3} as arising from multiple sheeted space-times. Such superposition attains a non-trivial meaning only when we consider measurement and expectation values of operators for which interference of the partial waves plays a physical role. Since we have not yet introduced these quantities we will consider it more carefully when it is necessary.

\subsection{Dirac equation solutions that are not eigenstates}

Ninth, when interpreted in light of ZM theory, the product decomposition of the Dirac spinor wavefunction in terms of a representation of the direction of time and a representation of the spatial variation of the clock-field allows the space of representations at a particular time to include states where the time direction is not aligned with the direction of maximum rate of change of the clock state. In this, the Dirac equation is like the ZM version of the classical variational principles, which begins by allowing the two directions to differ, but then sets the direction of time to be the direction of maximum rate of change of the clock. For example,  Eq. (\ref{generaltime}) can represent all of the possible directions of time, but a physical state is determined by a variational extremum. We have shown that once the direction of time is aligned with the direction of the maximum rate of change of the clock, ZM theory and the Dirac equation are in correspondence. This does not mean that when they are not aligned they correspond. The time direction is not aligned with the direction of maximum rate of change of the clock state for solutions of the Dirac equation that are formed as a superposition of eigenstates, but are not eigenstates. The dynamic behavior of the Dirac equation for non-eigenstates has therefore not been shown to be the same as that of the non-physical states discussed in ZM theory variation of the time direction. In particular, the dynamic behavior of non-eigenfunction solutions of the Dirac equation is determined by superposition of the dynamics of eigenstates, and therefore will be given meaning in the future discussion of superposition.

\subsection{Dirac's derivation}

Tenth, Dirac derived his equation from very few assumptions. The traditional derivation of Dirac's equation follows from:
\begin{itemize}
\item A differential equation relating time and space derivatives of a wavefunction representing a system.
\item A linear relationship between time and space derivatives.
\item Relativistic energy momentum relationship $\omega^2=k^2+m^2$ with $\omega$ and $k$ represented by time and space derivatives, and $m$ a constant.
\end{itemize}
These assumptions force the additional condition:
\begin{itemize}
\item Expansion to matrix coefficients to accommodate requirements of the coefficient algebra.
\end{itemize}
It appears that the assumptions of ZM theory are similar in requiring a particular relationship of the space and time derivatives. However, the difference resides in the self-consistency of the coefficients of the fundamental ZM theory Eq. (\ref{zmeq}), which are therefore dependent on the state of the system. For the case where the coefficients are assumed to be fixed, which is the case treated here, ZM theory and the Dirac equation give the same results. When we generalize the treatment of the ZM theory to the case of spatially varying coefficients, there is an inherent non-linearity which leads to coupling to an electromagnetic potential as will be discussed in a subsequent paper.

\subsection{Linear variations}

Eleventh, we have restricted our attention to linear variations of the field in space. For greater generality we would consider the environment to be incrementally specified in clock variation with each coordinate. Large scale observations would then be aggregates of small scale changes. It is interesting, however, that the linear treatment is insufficient, and we must incorporate ripples to be discussed in upcoming papers. Once the two different treatments are available, the incremental treatment will lead to a field theoretic formalism. The Dirac equation then provides linear components of the field theoretic treatment. 

\subsection{Relative representations}

Twelfth, conceptually there is a difference between the Dirac equation and ZM theory in the assignment of properties to system, environment or relationship between system and environment. The Dirac equation represents the velocity as a property of the particle. ZM theory considers it to a property of the relationship between the observer and the particle. It might be argued that in Dirac theory by performing a Lorentz transformation we see that the velocity is a property of the relationship between the observer and the particle. However, in the representation that is used, the velocity is contained in the wavefunction of the particle. In ZM theory the most natural equation is distinct in having the velocity as part of the equation (Hamiltonian operator) which is determined self-consistently from the state of the system.

\subsection{Toward considering measurement}

Thirteenth, the discussion of interpreting the Dirac spinor as a property of the coordinate system of the observer implies that there are changes in our understanding of the nature and paradoxes of measurement theory. Additional distinctions that are relevant have to do with the interpretation of superposition and the role of local causality. These issues will be further discussed in future papers.

\section{Summary}

In summary, this paper introduces a basic framework for discussing the correspondence of ZM theory with quantum mechanics, especially quantum electrodynamics. The dependence of the direction of time in ZM theory on the clock-field, and its non-orthogonality to space, changes the description of field properties from conventional equations that assume a fixed direction of time that is orthogonal to space. Correspondence is based upon showing that classes of clock-field variations can be described by traditional quantum and electrodynamic equations. The case of a constant rate of change of the clock-field in space was shown in this paper to correspond to the Dirac equation and the Klein-Gordon equation. For the Dirac equation, the wavefunction is a product of a plane wave representation of the clock-field, and a constant Dirac spinor that represents the transformation of the observer coordinate system as extrapolated into the observed region of space. The correspondence of the clock-field to the plane-wave representation suggests that we consider the clock phase as analogous to the phase of the quantum mechanical wave-function. The correspondence of the coordinate system extrapolation to the Dirac spinor describes the tipping of the direction of time as corresponding to a boost, and the rotation of the coordinate system as corresponding to a spinor rotation. Moreover, the inversion of the time axis and space axes correspond to the appearance of anti-particles, and spin degrees of freedom. A variety of comments point toward further theoretical developments. Specific steps still to be done include: (a) demonstrating that a variable rate of change of the clock-field in space, leading to a variable direction of time, yields contributions to the clock-field that correspond to electromagnetic fields as described by Maxwell's wave equation; (b) demonstrating the coupling of electromagnetic fields to particles; (c) describing new approaches to measurement theory; and (d) characterizing particle statistics.

I thank Marcus A. M. de Aguiar for helpful comments on the manuscript.

\end{document}